\RequirePackage{etoolbox}
\ifdef{\iftechreport}{}{\newif\iftechreport}
\documentclass[letterpaper,twocolumn,10pt]{article}
\usepackage{usenix}

\usepackage{tikz}
\usepackage{amsmath}

\usepackage{filecontents}
\usepackage{placeins}
\usepackage[export]{adjustbox}
\usepackage{url}
\usepackage{booktabs}
\usepackage{tabularx}
\usepackage{threeparttable}

\usepackage{siunitx}
\usepackage{multirow}
\usepackage{graphicx}
\usepackage{setspace}
\usepackage{float}
\usepackage{amsthm}
\usepackage{thmtools,thm-restate}
\declaretheoremstyle[
spaceabove=0pt,
spacebelow=0pt,
headindent=\parindent,
numbered=no,
headfont=\normalfont\bfseries,
bodyfont=\normalfont\itshape,
]{examplestyle}
\declaretheorem[style=examplestyle]{example}
\newtheorem{definition}{Definition}

\usepackage[ruled,vlined,linesnumbered]{algorithm2e}
\SetInd{0.5em}{0.3em}
\SetKwProg{Fn}{Function}{}{}
\SetVlineSkip{0pt}
\SetAlCapNameFnt{\small}
\SetAlCapFnt{\small}
\SetAlFnt{\small}
\makeatletter
\patchcmd\algocf@Vline{\vrule}{\vrule \kern-0.4pt}{}{}
\patchcmd\algocf@Vsline{\vrule}{\vrule \kern-0.4pt}{}{}
\newcommand{\removelatexerror}{\let\@latex@error\@gobble}
\makeatother
\usepackage{amsmath}
\usepackage{balance}
\usepackage{graphicx}
\usepackage[inline]{enumitem}
\setlist[itemize]{noitemsep,partopsep=0pt,topsep=0pt,leftmargin=1.5em}
\usepackage{mathtools}
\usepackage{verbatim}
\usepackage{threeparttable}
\usepackage{footnote}
\makesavenoteenv{tabular}
\makesavenoteenv{table}
\DeclarePairedDelimiter{\ceil}{\lceil}{\rceil}
\DeclarePairedDelimiter{\floor}{\lfloor}{\rfloor}

\usepackage[capitalise, noabbrev]{cleveref}
\usepackage{xcolor}

\newcommand{\eat}[1]{}

\apptocmd\normalsize{%
	\setlength{\abovedisplayskip}{1pt}
	\setlength{\belowdisplayskip}{1pt}
	\setlength{\abovedisplayshortskip}{1pt}
	\setlength{\belowdisplayshortskip}{1pt}
}{}{}

\newcommand{\ce}[1]{{\color{black}{#1}}}

\begin{document}
	
	\date{}
	
	\title{\Large \bf COLE: A Column-based Learned Storage for Blockchain Systems (Technical Report)}
	\author{
		{\rm Ce Zhang$^{*}$, Cheng Xu$^{*}$, Haibo Hu$^{\dagger}$, Jianliang Xu$^{*}$}\\
		$^{*}$Hong Kong Baptist University \, $^{\dagger}$Hong Kong Polytechnic University
	} 
	\maketitle
	\begin{abstract}
		Blockchain systems suffer from high storage costs as every node needs to store and maintain the entire blockchain data. After investigating Ethereum's storage, we find that the storage cost mostly comes from the index, i.e., Merkle Patricia Trie (MPT). \ce{To support provenance queries, MPT persists the index nodes during the data update, which adds too much storage overhead.} To reduce the storage size, an initial idea is to leverage the emerging learned index technique, which has been shown to have a smaller index size and more efficient query performance. However, directly applying it to the blockchain storage results in even higher overhead \ce{owing to the requirement of persisting index nodes} and the learned index's large node size. 
		To tackle this, we propose COLE, a novel column-based learned storage for blockchain systems. We follow the column-based database design to contiguously store each state's historical values, which are indexed by learned models to facilitate efficient data retrieval and provenance queries. We develop a series of write-optimized strategies to realize COLE in disk environments. Extensive experiments are conducted to validate the performance of the proposed COLE system. Compared with MPT, COLE reduces the storage size by up to 94\% while improving the system throughput by $1.4\times$-$5.4\times$.
	\end{abstract}
	
	\section{Introduction}\label{sec:intro}
	
	Blockchain, as the backbone of cryptocurrencies and decentralized applications~\cite{nakamoto2008bitcoin, wood2014ethereum}, is an immutable ledger built on a set of transactions agreed upon by untrusted nodes. It employs cryptographic hash chains and consensus protocols for data integrity. Users can retrieve historical data from blockchain nodes with integrity assurance, also known as provenance queries. However, \ce{all nodes are required to store the complete transactions and ledger states}, leading to amplified storage expenses, particularly as the blockchain continues to grow. \ce{For example, the Ethereum blockchain requires about 16TB storage as of December 2023, with an annual growth of around 4TB~\cite{ethsize}. This storage requirement may compel the resource-limited nodes to retain only the data of a few recent blocks, which restricts the ability to support data provenance. The nodes that maintain the complete data may also leave the network due to the rapidly increasing storage size, which potentially affects system security.} 
	
	\begin{figure}[t]
		\centering
		\includegraphics[width=0.85\linewidth]{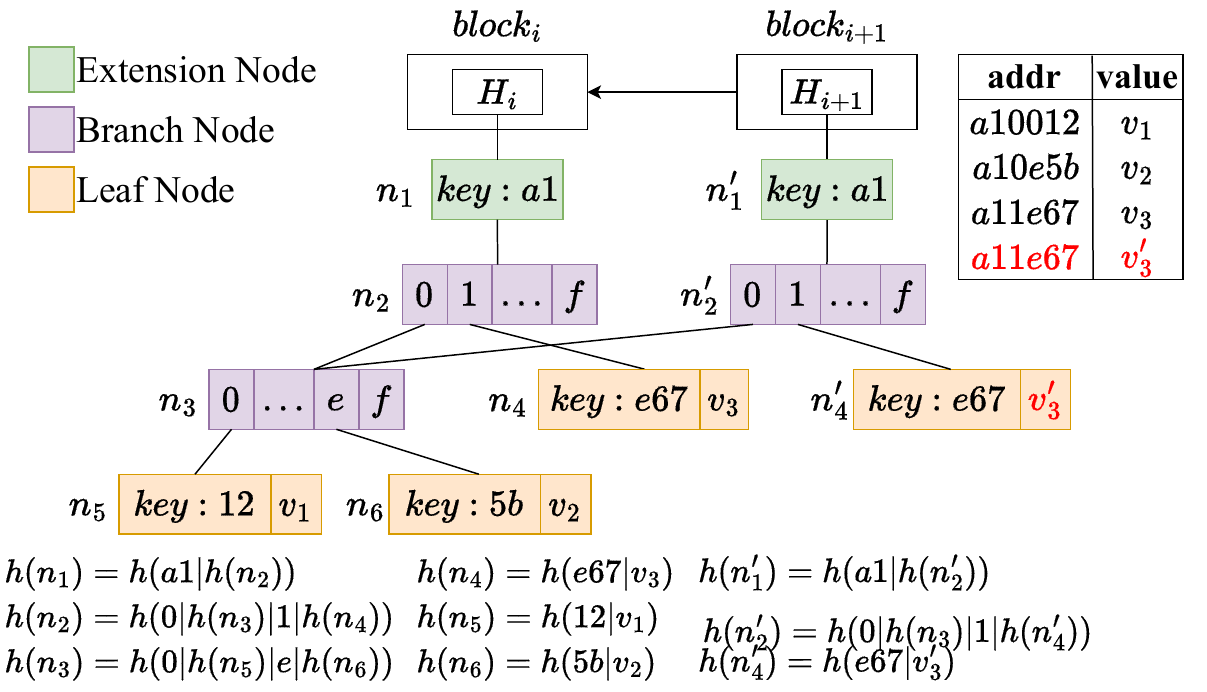}
		\vspace{-0.4em}
		\caption{An Example of Merkle Patricia Trie}\label{fig:mpt}
	\end{figure}
	
	To tackle the storage issue, we investigate Ethereum's index, Merkle Patricia Trie (MPT), to identify the storage bottleneck. MPT combines Patricia Trie with Merkle Hash Tree (MHT)~\cite{merkle1989certified} to ensure data integrity. \ce{During data updates, its index nodes are persisted to support provenance queries. \Cref{fig:mpt} shows an example of an MPT storing three state addresses across two blocks.}
	Each node is augmented with a digest from its content and child nodes (e.g., $h(n_1)=h(a1|h(n_2))$). The root hash secures data integrity through the collision-resistance of the cryptographic hash function and the hierarchical structure. With each new block, MPT retains obsolete nodes from the preceding block. For example, in block $i+1$, updating address $a11e67$ with $v_3'$ introduces new nodes $n_1', n_2', n_4'$, while old nodes $n_1, n_2, n_4$ endure. This setup allows historical data retrieval from any block (e.g., for address $a11e67$ in block $i$, value $v_3$ is retrieved by traversing nodes $n_1$, $ n_2$, and $n_4$).
	
	However, this approach adds too much storage overhead due to duplicating nodes along the update path (e.g., $n_1, n_2, n_4$ and $n_1', n_2', n_4'$ in \cref{fig:mpt}). Consequently, most storage overhead comes from the index rather than the underlying data. \ce{In a preliminary experiment with 10 million transactions under the SmallBank workload~\cite{dinh2017blockbench}, we observed that the underlying data contributes only $2.8\%$ of the total storage.} Thus, a more compact index supporting data integrity and provenance queries is imperative.  
	
	Recently, a novel indexing technique, learned index~\cite{kraska2018case, ferragina2020pgm, ding2020alex, wu2021updatable}, has emerged and shows notably smaller index size and faster query speed. The improved performance comes from the substitution of the directing keys in index nodes with a learned model. For instance, consider a key-value database with linear key distribution: ${(1, v_1), (2, v_2), \cdots, (n, v_{n})}$. In a traditional B+-tree with fanout $f$, this leads to $O(\frac{n}{f})$ nodes and $O(\log_f n)$ levels, resulting in $O(n)$ storage costs and $O(\log_f n \cdot \log_2 f)$ query times. Conversely, using a simple linear model $y=x$ enables accurate data positioning with just $O(1)$ storage and $O(1)$ query times. Although this example may not perfectly reflect real-world applications, it highlights that the learned index outperforms traditional indexes significantly when the model effectively learns the data.
	
	In view of the advantages of the learned index, one may want to apply it to blockchain storage to improve performance. However, the current learned indexes do not support both data integrity and provenance queries required by blockchain systems. A naive approach is to combine the learned index with MHT~\cite{merkle1989certified} and make the index nodes persistent, as in MPT. Nonetheless, this is not feasible due to the larger node size of the learned index. The fanout of such a node is mainly dictated by data distribution. In favorable cases, only a few models are needed to index data, leading to a node fanout comparable to data magnitude. Thus, persisting learned index nodes might incur even higher storage overhead than MPT. Our evaluation in \cref{sec:evaluation} shows that a learned index with persistent nodes is $5\times$ to $31\times$ larger than MPT. Furthermore, as blockchain systems require durable disk-based storage and often involve frequent data updates, the learned index should be optimized for both disk and write operations. 
	Therefore, a blockchain-friendly learned index needs to be proposed.
	
	In this paper, we propose COLE, a novel \underline{co}lumn-based \underline{le}arned storage for blockchain systems that overcomes the limitations of current learned indexes and supports provenance queries. The key challenge in adapting learned indexes to blockchains is the need for node persistence, which may lead to substantial storage overhead. COLE tackles this issue with an innovative \emph{column-based} design, inspired by column-based databases~\cite{abadi2009column, mehra2015column}. In this design, each ledger state is treated as a ``column'', with different versions of a state stored contiguously and indexed using \emph{learned models} within \emph{the latest block's} index. This enables efficient data updates as append operations with associated version numbers (i.e., state's block heights). Moreover, historical data queries no longer traverse previous block indexes, but utilize the learned index in the most recent block. The column-based design also simplifies model learning and reduces disk IOs.
	
	To handle frequent data updates and enhance write efficiency in COLE, we propose adopting the \emph{log-structured merge-tree} (LSM-tree)~\cite{o1996log, li2009tree} maintenance approach to manage the learned models. This involves inserting updates into an in-memory index before merging them into on-disk levels that grow exponentially. For each on-disk level, we design a disk-optimized learned model that can be constructed in a \emph{streaming} way, which enables efficient data retrieval with minimal IO cost. To guarantee data integrity, we construct an $m$-ary complete MHT for the blockchain data in each on-disk level. The root hashes of the in-memory index and all MHTs combine to create a root digest that attests to the entire blockchain data. However, recursive merges during write operations can lead to long-tail latency in the LSM-tree approach. To alleviate this issue, we further develop a novel checkpoint-based asynchronous merge strategy to ensure the synchronization of the storage among blockchain nodes.
	
	To summarize, this paper makes the following contributions:
	\begin{itemize}
		\item To the best of our knowledge, COLE is the first column-based learned storage that combines learned models with the column-based design to reduce storage costs for blockchain systems. 
		\item We propose novel write-optimized and disk-optimized designs to store blockchain data, learned models, and Merkle files for realizing COLE.
		\item We develop a new checkpoint-based asynchronous merge strategy to address the long-tail latency problem for data writes in COLE.
		\item We conduct extensive experiments to evaluate COLE's performance. The results show that compared with MPT, COLE reduces storage size by up to 94\% and improves system throughput by $1.4\times$-$5.4\times$. Additionally, the proposed asynchronous merge decreases long-tail latency by 1-2 orders of magnitude while maintaining a comparable storage size.
	\end{itemize}
	
	The rest of the paper is organized as follows. We present some preliminaries about blockchain storage in \cref{sec:prelim-blockchain-index}. \Cref{sec:overview} gives a system overview of COLE.
	\Cref{sec:write-operation} designs the write operation of COLE, followed by an asynchronous merge strategy in
	\Cref{sec:async-write}. \Cref{sec:read-operation} describes the read operations of COLE. \Cref{sec:analysis} presents a complexity analysis. The experimental evaluation results are shown in \cref{sec:evaluation}. \Cref{sec:relatedworks} discusses the related work. Finally, we conclude our paper in \cref{sec:conclusion}.
	
	\section{Blockchain Storage Basics}\label{sec:prelim-blockchain-index}
\begin{figure}[t]
		\centering
		\includegraphics[width=0.8\linewidth]{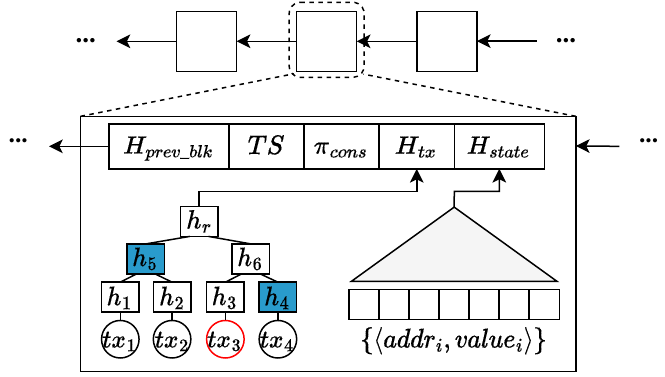}
		\vspace{-0.4em}
		\caption{Block Data Structure}\label{fig:block-structure}
	\end{figure}

	In this section, we give some necessary preliminaries to introduce the proposed COLE. Blockchain is a chain of blocks that maintains a set of states and records the transactions that modify these states. \ce{To establish a consistent view of the states among mutually untrusted blockchain nodes, a consensus protocol is utilized to globally order the transactions~\cite{nakamoto2008bitcoin, saleh2018blockchain, castro2002practical}.} The transaction's execution program is known as \emph{smart contract}. A smart contract can store states, each of which is identified by a state address $addr$. In Ethereum~\cite{wood2014ethereum}, both the state address $addr$ and the state value $value$ are fixed-sized strings. \Cref{fig:block-structure} shows an example of the block data structure. The header of a block consists of (i) $H_{prev\_blk}$, the hash of the previous block; (ii) $TS$, the timestamp; (iii) $\pi_{cons}$, the consensus protocol related data; (iv) $H_{tx}$, the root digest of the transactions in the current block; (v) $H_{state}$, the root digest of the states. The block body includes the transactions, states, and their corresponding Merkle Hash Tree (MHTs).
	
	MHT is a prevalent hierarchical structure to ensure data integrity~\cite{merkle1989certified}. \ce{In the context of blockchain, MHT is built for the transactions of each block and the ledger states. \Cref{fig:block-structure} shows an example of an MHT of a block's transactions. The leaf nodes are the hash values of the transactions (e.g., $h_1=h(tx_1)$). The internal nodes are the hash values of their child nodes (e.g., $h_5=h(h_1||h_2)$). MHT enables the proof of existence for a given transaction. For example, to prove $tx_3$, the sibling hashes along the search path (i.e., $h_4$ and $h_5$, shaded in \cref{fig:block-structure}) are returned as the proof.} One can verify $tx_3$ by reconstructing the root hash using the proof (i.e., $h(h_5||h(h(tx_3)||h_4))$) and comparing it with the one in the block header (i.e., $H_{tx}$). Apart from being used in the blockchain, MHT has also been extended to database indexes to support result integrity verification for different queries. For example, MHT has been extended to Merkle B+-tree (MB-tree) by combining the Merkle structure with B+-tree, to support trustworthy queries in relational databases~\cite{li2006dynamic}.
	
	The blockchain storage uses an index to efficiently maintain and access the states~\cite{wood2014ethereum, wang2018forkbase}. Besides the write and read operations that a normal index supports, the index of the blockchain storage should also fulfill the two requirements we mentioned before: (i) ensuring the \emph{integrity} of the indexed blockchain states, (ii) supporting \emph{provenance queries} that enable blockchain users to retrieve historical state values with integrity assurance. With these requirements, the index of the blockchain storage should support the following functions:
	\begin{itemize}
		\item $\FuncSty{Put}(addr, value)$: insert the state with the address $addr$ and the value $value$ to the current block;
		\item $\FuncSty{Get}(addr)$: return the \emph{latest} value of the state at address $addr$ if it exists, or returns $nil$ otherwise;
		\item $\FuncSty{ProvQuery}(addr, [blk_l, blk_u])$: return the provenance query results $\{value\}$ and a proof $\pi$, given the address $addr$ and the block height range $[blk_l, blk_u]$;
		\item $\FuncSty{VerifyProv}(addr, [blk_l, blk_u], \{value\}, \pi, H_{state})$: verify the \allowbreak provenance query results $\{value\}$ w.r.t.~the address, the block height range, the proof, and $H_{state}$, where $H_{state}$ is the root digest of the states.
	\end{itemize}
	
	Ethereum employs Merkle Patricia Trie (MPT) to index blockchain states. In \cref{sec:intro}, we have shown how MPT implements $\FuncSty{Put}(\cdot)$ and $\FuncSty{ProvQuery}(\cdot)$ using \cref{fig:mpt} and the address $a11e67$. We now explain the other two functions using the same example. $\FuncSty{Get}(a11e67)$ finds $a11e67$'s latest value $v_3'$ by traversing $n_1', n_2', n_4'$ under the latest block $i+1$. After $\FuncSty{ProvQuery}(a11e67, [i,i])$ gets $v_3$ and the proof $\pi = \{n_1, n_2, n_4, h(n_3)\}$ in block $i$, $\FuncSty{VerifyProv}(\cdot)$ is used to verify the integrity of $v_3$ by reconstructing the root digest using the nodes from $n_4$ to $n_1$ in $\pi$ and checks whether the reconstructed one matches the public digest $H_i$ in block $i$ and whether the search path in $\pi$ corresponds to the address $a11e67$.
	
	\vspace{-0.4em}
	\section{COLE Overview}\label{sec:overview}
	This section presents COLE, our proposed column-based learned storage for blockchain systems. We first give the design goals and then show how COLE achieves these goals.
	
	\vspace{-0.4em}
	\subsection{Design Goals}
	We aim to achieve the following design goals for COLE:
	\begin{itemize}
		\item \textbf{Minimizing storage size.} To scale up the blockchain system, it is important to reduce the storage size by leveraging the learned index and column-based design.
		\item \textbf{Supporting the requirements of blockchain storage.} As blockchain storage, it should ensure data integrity and support provenance queries as mentioned in \cref{sec:prelim-blockchain-index}.
		\item \textbf{Achieving efficient writes in a disk environment.}
		Since blockchain is write-intensive and all data needs to be preserved on disk, the system should be write-optimized and disk-optimized to achieve better performance.
	\end{itemize}
	
	\vspace{-0.4em}
	\subsection{Design Overview}\label{sec:design-overview}
	
	\begin{figure}[t]
		\centering
		\includegraphics[width=.85\linewidth]{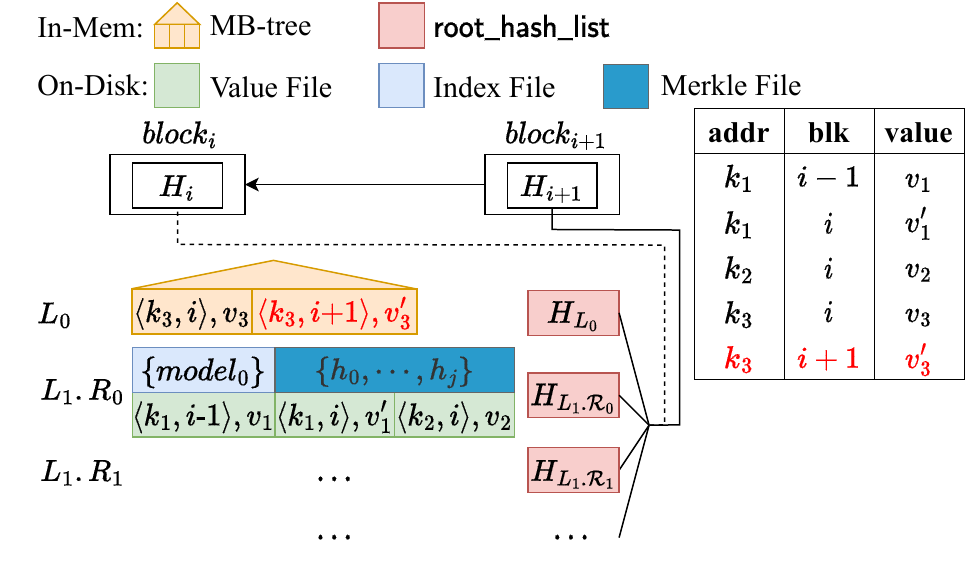}
		\vspace{-0.4em}
		\caption{Overview of COLE}\label{fig:overview}
	\end{figure}
	
	\Cref{fig:overview} shows the overview of COLE. Following the column-based design~\cite{abadi2009column, mehra2015column}, we adopt an analogy between blockchain states and database columns. Each state's historical versions are contiguously stored in the index of the latest block. \ce{When a state is updated in a new block, the state and its version number (i.e., block height) are appended to the index where all of the state's historical versions are stored.} For indexing historical state values, we use a \emph{compound key} $\mathcal{K}$ in the form of $\langle addr, blk\rangle$, where $blk$ is the block height when the value of $addr$ was updated. \ce{In \cref{fig:overview}, when block $i+1$ updates the state at address $k_3$ (highlighted in red), a new compound key of $k_3$, $\mathcal{K}_3' \gets \langle k_3, i+1\rangle$, is created. Then, the updated value $v_3'$ indexed by $\mathcal{K}_3'$ is inserted into COLE. With the column-based design, $v_3'$ is stored next to $k_3$'s old version $v_3$.} Compared with the MPT in \cref{fig:mpt}, the cumbersome node duplication along the update path (e.g., $n_1, n_2, n_4$ and $n_1', n_2', n_4'$) is avoided to save the storage overhead.

	To mitigate the high write cost associated with learned models for indexing blockchain data in a column-based design, we propose using the LSM-tree maintenance strategy in COLE. It structures index storage into levels of exponentially increasing sizes. New data is initially added to the first level. \ce{When the level reaches its pre-defined maximum capacity, all the data in that level is merged into a sorted run in the next level. This merge operation can occur recursively until the capacity requirement is no longer violated.} The first level, often highly dynamic, is typically stored in memory, while other levels reside on disk. COLE employs Merkle B+-tree (MB-tree)~\cite{li2006dynamic} for the first level and disk-optimized learned indexes for subsequent levels. We choose MB-tree over MPT for the in-memory level due to its better efficiency in compacting data into sorted runs and flushing them to the first on-disk level. 
	
	Each on-disk level contains a fixed number of sorted runs, each of which is associated with a value file, an index file, and a Merkle file:
	\begin{itemize}
		\item \textbf{Value file} stores blockchain states as compound key-value pairs, which are ordered by their compound keys to facilitate the learned index.
		\item \textbf{Index file} helps locate blockchain states in the value file during read operations. It uses a disk-optimized learned index, inspired by PGM-index~\cite{ferragina2020pgm}, for efficient data retrieval with minimal IO cost.
		\item \textbf{Merkle file} authenticates the data stored in the value file. It is an $m$-ary complete MHT built on the compound key-value pairs.
	\end{itemize}
	\ce{Note that since the model construction and utilization require numerical data types, we convert a compound key into a big integer using the binary representation of the address and the block height. For example, given a compound key $\mathcal{K}\gets\langle addr, blk\rangle$, its big integer is computed as $binary(addr)\times 2^{64}+blk$, where $blk$ is a 64-bit value.} Moreover, to ensure data integrity, root hashes of both the in-memory MB-tree and the Merkle files of each on-disk run are combined to create a \textsf{root\_hash\_list}. The root digest of states, stored in the block header, is computed from this list. This list is cached in memory to expedite root digest computation.

	With this design, to retrieve the state value of address $addr_q$ at a block height $blk_q$, a compound key $\mathcal{K}_q \gets \langle addr_q, blk_q \rangle$ is employed. The process entails a level-wise search within COLE, initiated from the first level. The MB-tree or the learned indexes in other levels are traversed. The search ceases upon encountering a compound key $\mathcal{K}_r \gets \langle addr_r, blk_r \rangle$ where $addr_r = addr_q$ and $blk_r \le blk_q$, at which point the corresponding value is returned. For retrieving the latest value of a state, the procedure remains similar but with the search key set to $\langle addr_q, max\_int \rangle$, where $max\_int$ is the maximum integer. That is, the search is stopped as long as a state value with the queried address $addr_q$ is found.
	
	\section{Write Operation of COLE}\label{sec:write-operation}
	\begin{figure}
		\removelatexerror
		\begin{algorithm}[H]
			\caption{Write Algorithm}\label{alg:complete-insertion}
			\SetKwFunction{FMain}{Put}
			\Fn{\FMain{$addr, value$}}{
				\KwIn{State address $addr$, value $value$}
				$blk\gets$ current block height;\label{sec:write-operation-1}
				$\mathcal{K}\gets \langle addr, blk\rangle$\;
				
				Insert $\langle \mathcal{K}, value\rangle$ into the MB-tree in $L_0$\;\label{sec:write-operation-2}
				\If{$L_0$ contains $B$ compound key-value pairs}{
					Flush the leaf nodes in $L_0$ to $L_1$ as a sorted run\;\label{sec:write-operation-3}
					Generate files $\mathcal{F}_{V}, \mathcal{F}_{I}, \mathcal{F}_{H}$ for this run\;\label{sec:write-operation-4}
					$i \gets 1$\;
					\While{$L_i$ contains $T$ runs}{\label{sec:write-operation-5}
						Sort-merge all the runs in $L_i$ to $L_{i+1}$ as a new run\;
						Generate files $\mathcal{F}_{V}, \mathcal{F}_{I}, \mathcal{F}_{H}$ for the new run\;
						Remove all the runs in $L_i$\;\label{sec:write-operation-6}
						$i \gets i+1$\;
					}
				}
				Update $H_{state}$ when finalizing the current block\;\label{sec:write-operation-7}
			}
		\end{algorithm}
	\end{figure}
	
	We now detail the write operation of COLE.
	\ce{As mentioned in \cref{sec:design-overview}, COLE organizes the storage using an LSM-tree, which consists of an in-memory level and multiple on-disk levels. The in-memory level has a capacity of $B$ states in the form of compound key-value pairs. Once this capacity is reached, the in-memory level is flushed to the disk as a sorted run. Similarly, when the first on-disk level reaches its capacity of $T$ sorted runs, they are merged into a new run in the next level. This merging process continues for subsequent disk levels, with the size of each run growing exponentially with a ratio of $T$. That is, level $i$ has a maximum of $B \cdot T^i$ states.}
	
	
	\Cref{alg:complete-insertion} shows COLE's write operation. It starts by calculating a compound key for the state using the address and the current block height (\cref{sec:write-operation-1}). The compound key-value pair is inserted into the in-memory level $L_0$ indexed by the MB-tree (\cref{sec:write-operation-2}). As $L_0$ fills up, it is flushed to the first on-disk level $L_1$ as a sorted run (\cref{sec:write-operation-3}). The value file $\mathcal{F}_V$ is generated by scanning compound key-value pairs in the MB-tree's leaf nodes (\cref{sec:write-operation-4}). At the same time, the index file $\mathcal{F}_I$ and the Merkle file $\mathcal{F}_H$ are constructed in a \emph{streaming} manner (see \cref{sec:model-file}, \cref{sec:merkle-file} for details). When on-disk level $L_i$ fills up (i.e., with $T$ runs), all the runs in $L_i$ are merge-sorted as a new run in the next level $L_{i+1}$, with corresponding three files generated (\crefrange{sec:write-operation-5}{sec:write-operation-6}). This level-merge process continues recursively until a level does not fill up. The blockchain's state root digest $H_{state}$ is computed by hashing the concatenation of the root hash of $L_0$'s MB-tree and root hashes of runs in other levels, stored in \textsf{root\_hash\_list}, when finalizing the current block (\cref{sec:write-operation-7}).

	\begin{figure}[t]
		\centering
		\includegraphics[width=0.85\linewidth]{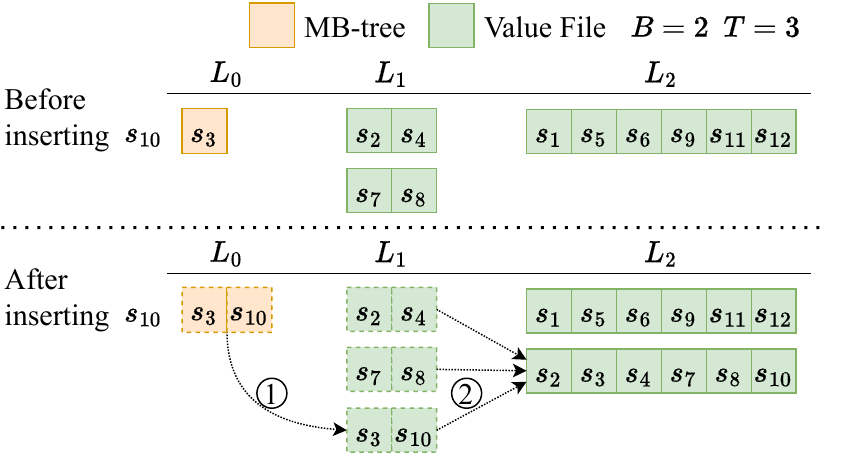}
		\vspace{-0.4em}
		\caption{An Example of Write Operation}\label{fig:level-merge}
	\end{figure}
	
	\begin{example}
		\Cref{fig:level-merge} shows an example of the insertion of $s_{10}$. For clarity, we show only the states and the value files but omit the index files and Merkle files. Assume $B=2$ and $T=3$. The sizes of the runs in $L_1$ and $L_2$ are $2$ and $6$, respectively. After $s_{10}$ is inserted into in-memory level $L_0$, the level is full and its states are flushed to $L_1$ as a sorted run (step \raisebox{.5pt}{\textcircled{\raisebox{-.9pt} {1}}}). This incurs $L_1$ reaching the maximum number of runs. Thus, all the runs in $L_1$ are next sort-merged as a new run, placed in $L_2$ (step \raisebox{.5pt}{\textcircled{\raisebox{-.9pt} {2}}}). Finally, $L_0$ and $L_1$ are empty and $L_2$ has two runs, each of which contains six states.
	\end{example}
	
	\ce{A common optimization technique to speed up read operations is to integrate a Bloom filter into the in-memory MB-tree and each run in the on-disk levels. We incorporate the Bloom filter into COLE with careful consideration. First, they should be built upon the addresses of the underlying states rather than their compound keys to facilitate read operations. Second, since the Bloom filters may produce false positives, if they indicate that an address exists, we further resort to the normal read process of the corresponding MB-tree or the disk run to ensure the search correctness. We will elaborate on their usage during the read operation in \cref{sec:read-operation}. Moreover, the Bloom filters should be incorporated alongside the root hashes of each run when computing the states' root digest. This is needed to verify the result integrity during provenance queries.}
	
	
	\subsection{Index File Construction}\label{sec:model-file}
	An index file consists of the models that can be used to locate the positions of the states' compound keys in the value file. Inspired by PGM-index~\cite{ferragina2020pgm}, we start by defining an $\epsilon$-bounded piecewise linear model (or \emph{model} for short) as follows.
	\begin{definition}[$\epsilon$-Bounded Piecewise Linear Model]
		The model is a tuple of $\mathcal{M} = \langle sl, ic, k_{min}, p_{max} \rangle$, where $sl$ and $ic$ are the slope and intercept of the linear model, $k_{min}$ is the first key in the model, and $p_{max}$ is the last position of the data covered by the model.
	\end{definition}
	Given a model, one can predict a compound key $\mathcal{K}$'s position $p_{real}$ in a file, if $\mathcal{K} \ge k_{min}$. The predicted position $p_{pred}$ is calculated as $p_{pred} = \min(\mathcal{K} \cdot sl + ic, p_{max})$, which satisfies $|p_{pred} - p_{real}| \le \epsilon$. \ce{Since files are often organized into pages, we set $\epsilon$ as half the number of models that can fit into a single disk page to generate the models in a disk-friendly manner.} As will be shown, this reduces the IO cost by ensuring that at most two pages need to be accessed per model during read operations.
	
	\begin{algorithm}[t]
		\caption{Learn Models from a Stream}\label{alg:model-construction}
		\SetKwFunction{FMain}{BuildModel}
		\Fn{\FMain{$\mathcal{S}, \epsilon$}}{
			\KwIn{Input stream $\mathcal{S}$, error bound $\epsilon$}
			\KwOut{A stream of models $\{\mathcal{M}\}$}
			$k_{min} \gets \emptyset$, $p_{max} \gets \emptyset$, $g_{last} \gets \emptyset$\;
			Init an empty convex hull $\mathcal{H}$\;
			\ForEach{$\langle \mathcal{K}, p_{real} \rangle \gets \mathcal{S}$}{
				\lIf{$k_{min} = \emptyset$}{$k_{min} \gets \mathcal{K}$}
				Add $\langle BigNum(\mathcal{K}), p_{real} \rangle$ to $\mathcal{H}$\;
				Compute the minimum parallelogram $\mathcal{G}$ that covers $\mathcal{H}$\;
				\eIf{$\mathcal{G}.height \le 2 \epsilon$}{
					$p_{max}\gets p_{real}, g_{last} \gets \mathcal{G}$\;
				}{
					Compute slope $sl$ and intercept $ic$ from $g_{last}$\;
					$\mathcal{M} \gets \langle sl, ic, k_{min}, p_{max} \rangle$\;
					\textbf{yield} $\mathcal{M}$\;
					
					$k_{min} \gets \mathcal{K}$\;
					Init a new convex hull $\mathcal{H}$ with $\langle BigNum(\mathcal{K}), p_{real} \rangle$\;
				}
			}
		}
	\end{algorithm}
	To compute models from a stream of compound keys and their corresponding positions, we treat each compound key and its position as a point's coordinates. \ce{Upon the arrival of a new compound key, we convert it into a big integer using the binary representation of the address and the block height as mentioned in \cref{sec:design-overview}. Next, we find the smallest convex shape containing all the existing input points, which is known as a convex hull. Note that this convex hull can be computed incrementally in a streaming fashion \cite{o1981line}.} Then, we find the minimal parallelogram that covers the convex hull, with one side aligned to the vertical axis (i.e., the position axis). If the parallelogram's height stays under $2\epsilon$, all existing inputs can fit into a single model. In this case, we try to include the next compound key in the stream for model construction. However, if the current parallelogram fails to meet the height criteria, the slope and intercept of the central line in the parallelogram will be used to build a model that covers all existing compound keys except the current one. After this, a new model will be built, starting from the current compound key. We summarize the algorithm in \cref{alg:model-construction}.
	
	\begin{figure}[t]
		\centering
		\includegraphics[width=0.85\linewidth]{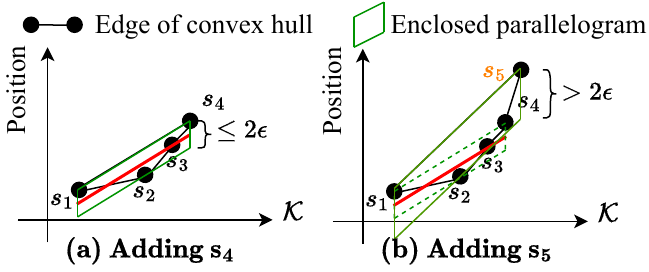}
		\vspace{-0.5em}
		\caption{An Example of Model Learning}\label{fig:pgm-example}
	\end{figure}
	\begin{example}
		\ce{\Cref{fig:pgm-example} shows an example of model learning from a stream. Assume states $s_1$ to $s_3$ form a convex hull, with its minimal parallelogram satisfying the height criterion (i.e., below $2\epsilon$). After state $s_4$ is added, the parallelogram's height remains within $2\epsilon$ (see \cref{fig:pgm-example}(a)), indicating that states $s_1$ to $s_4$ can be fit into one model. However, after the next state $s_5$ is added, the parallelogram's height exceeds $2\epsilon$ (see \cref{fig:pgm-example}(b)). Thus, the slope and intercept of the previous parallelogram's central line (highlighted in red) are used to build a model for $s_1$ to $s_4$, with $s_5$ reserved for the next model.}
	\end{example}
	
	\begin{algorithm}[t]
		\caption{Index File Construction}\label{alg:model-file-construction}
		\SetKwFunction{FMain}{ConstructIndexFile}
		\Fn{\FMain{$\mathcal{S}, \epsilon$}}{
			\KwIn{Input stream $\mathcal{S}$ of compound key-position pairs}
			\KwOut{Index file $\mathcal{F}_I$}
			Create an empty index file $\mathcal{F}_I$\;
			Invoke $\FuncSty{BuildModel}(\mathcal{S}, \epsilon)$ and write to $\mathcal{F}_I$\;\label{alg:model-file-construction-1}
			$n \gets \text{\# of pages in }\mathcal{F}_I$\;\label{alg:model-file-construction-2}
			\While{$n > 1$}{
				$\mathcal{S} \gets \{ \langle \mathcal{M}.k_{min}, pos \rangle~|~\text{\textbf{foreach}}~\langle \mathcal{M}, pos \rangle \in \mathcal{F}_I[-n:] \}$\; \label{alg:model-file-construction-4}
				Invoke $\FuncSty{BuildModel}(\mathcal{S}, \epsilon)$ and append to $\mathcal{F}_I$\;\label{alg:model-file-construction-5}
				$n \gets \text{\# of pages in }\mathcal{F}_I - n$\;\label{alg:model-file-construction-3}
			}
			\Return{$\mathcal{F}_I$}\;
		}
	\end{algorithm}
	
	\Cref{alg:model-file-construction} shows the overall procedure of index file generation. During flush or sort-merge operations in \cref{alg:complete-insertion}, ordered compound keys and state values are generated and written streamingly into the value file. Meanwhile, another stream consisting of compound keys and their positions is created and used to generate models with \cref{alg:model-construction} (\cref{alg:model-file-construction-1}). Once the models are yielded by \cref{alg:model-construction}, they are immediately written to the index file, constituting the bottom layer of the run's learned index. Then, we recursively build the upper layers of the index until the top layer can fit into a single disk page (\crefrange{alg:model-file-construction-2}{alg:model-file-construction-3}). \ce{Specifically, for each layer, we scan lower-layer models (denoted as $\mathcal{F}_I[-n:]$) to create a compound key stream using $k_{min}$ in each model and their index file positions (\cref{alg:model-file-construction-4}).} Similar to the bottom layer, we use \cref{alg:model-construction} on the stream to create models and instantly write them to the index file (\cref{alg:model-file-construction-5}). This results in the sequential storage of models across layers in a bottom-up manner. \ce{The index file remains valid from its construction until the next level merge operation thanks to the LSM-tree-based maintenance approach, which avoids costly model retraining.}

	\subsection{Merkle File Construction}\label{sec:merkle-file}
	
	\begin{algorithm}[t]
		\caption{Merkle File Construction}\label{alg:merkle-file-construction}
		\SetKwFunction{FMain}{ConstructMerkleFile}
		\Fn{\FMain{$\mathcal{S}, n, m$}}{
			\KwIn{Input stream $\mathcal{S}$ of compound key-value pairs, stream size $n$, fanout $m$}
			\KwOut{Merkle file $\mathcal{F}_H$}
			\ce{$N_{nodes} \gets [n, \ceil{\frac{n}{m}}, \ceil{\frac{n}{m^2}}, \cdots, 1]$, $d \gets |N_{nodes}|$\;}\label{alg:merkle-file-construction-1}
			$\text{layer\_offset}[0] \gets 0$\;\label{alg:merkle-file-construction-2}
			$\text{layer\_offset}[i] \gets \sum_{0}^{i-1} N_{nodes}[i-1]$, $\forall i \in [1, d-1]$\;\label{alg:merkle-file-construction-3}
			Create a merkle file $\mathcal{F}_H$ with size $\sum_{i=0}^{d-1} N_{nodes}[i]$\;
			Create a cache $\mathcal{C}$ with $d$ number of buffers\;
			\ForEach{$\langle \mathcal{K}, value \rangle \gets S$}{
				$h' \gets h(\mathcal{K} \| value)$, append $h'$ to $\mathcal{C}[0]$\;\label{alg:merkle-file-construction-4}
				\ForEach{$i$~\textbf{in}~$0$~\textbf{to}~$d - 2$}{
					\uIf{$|\mathcal{C}[i]| = m$}{
						$h' \gets h(\mathcal{C}[i])$, append $h'$ to $\mathcal{C}[i + 1]$\;\label{alg:merkle-file-construction-5}
						Flush $\mathcal{C}[i]$ to $\mathcal{F}_H$ at offset $\text{layer\_offset}[i]$\;\label{alg:merkle-file-construction-6}
						$\text{layer\_offset}[i] \gets \text{layer\_offset}[i] + m$\;\label{alg:merkle-file-construction-7}
					}
					\lElse{\textbf{break}}
				}
			}
			\ForEach{$i$~\textbf{in}~$0$~\textbf{to}~$d-1$}{\label{alg:merkle-file-construction-8}
				\If{$\mathcal{C}[i]$ is not empty}{
					$h' \gets h(\mathcal{C}[i])$, append $h'$ to $\mathcal{C}[i + 1]$\;
					Flush $\mathcal{C}[i]$ to $\mathcal{F}_H$ at offset $\text{layer\_offset}[i]$\;\label{alg:merkle-file-construction-9}
				}
			}
			\Return{$\mathcal{F}_H$}\;
		}
	\end{algorithm}
	\ce{A Merkle file stores an $m$-ary complete MHT that authenticates the compound key-value pairs in the corresponding value file.} The related index file's learned models are excluded from authentication, as they solely enhance query efficiency and do not affect blockchain data integrity. For the $m$-ary complete MHT, the bottom layer consists of hash values of every compound key-value pair in the value file. The hash values in an upper layer are recursively computed from every $m$ hash values in the lower layer, except that the last one might be computed from less than $m$ hash values in the lower layer.
	
	\begin{definition}[Hash Value]
		A hash value in the bottom layer of the MHT is computed as $h_i = h(\mathcal{K}_i \| value_i)$, where $\mathcal{K}_i, value_i$ are the corresponding compound key and value, $\|$ is the concatenation operator, and $h(\cdot)$ is a cryptographic hash function such as SHA-256. A hash value in an upper layer of the MHT is computed as $h_i = h(h_i^1 \| h_i^2 \| \cdots \| h_i^{m^*})$, where $m^* \le m$ and $h_i^j$ is the corresponding $j$-th hash in the lower layer.
	\end{definition}
	
	Similar to Algorithm~\ref{alg:model-file-construction}, we streamingly generate the Merkle file. However, instead of layer-wise construction, we concurrently build all MHT layers to reduce IO costs, as shown in \cref{alg:merkle-file-construction}. Note that the size of the input stream of compound key-value pairs $n$ is known in advance since the size of a value file is determined by the level of its corresponding run. Thus, the MHT has $\ceil{\log_m n} +1$ layers, containing $n, \ceil{\frac{n}{m}}, \ceil{\frac{n}{m^2}}, \cdots, 1$ hash values (\cref{alg:merkle-file-construction-1}). Layer offsets can also be computed (\crefrange{alg:merkle-file-construction-2}{alg:merkle-file-construction-3}). For concurrent construction, $\ceil{\log_m n} +1$ buffers are maintained, one per layer. Upon the arrival of a new compound key-value pair, its hash value is computed and added to the bottom layer's buffer (\cref{alg:merkle-file-construction-4}). When a buffer fills with $m$ hash values, an upper layer's hash value is created and added to its buffer (\cref{alg:merkle-file-construction-5}). Next, the buffered hash values in the current layer are flushed to the Merkle file, followed by incrementing the offset (\crefrange{alg:merkle-file-construction-6}{alg:merkle-file-construction-7}). This process recurs in upper layers until a layer with less than $m$ buffered hash values is encountered. Once the input stream is fully processed, any remaining non-empty buffers will hold fewer than $m$ hash values. If so, we'll initiate this process by taking a buffer from the lowest layer and iteratively generating hash values. Each hash value is added to the upper layer before flushing the buffer to the Merkle file (\crefrange{alg:merkle-file-construction-8}{alg:merkle-file-construction-9}).

	\begin{figure}[t]
		\centering
		\includegraphics[width=0.85\linewidth]{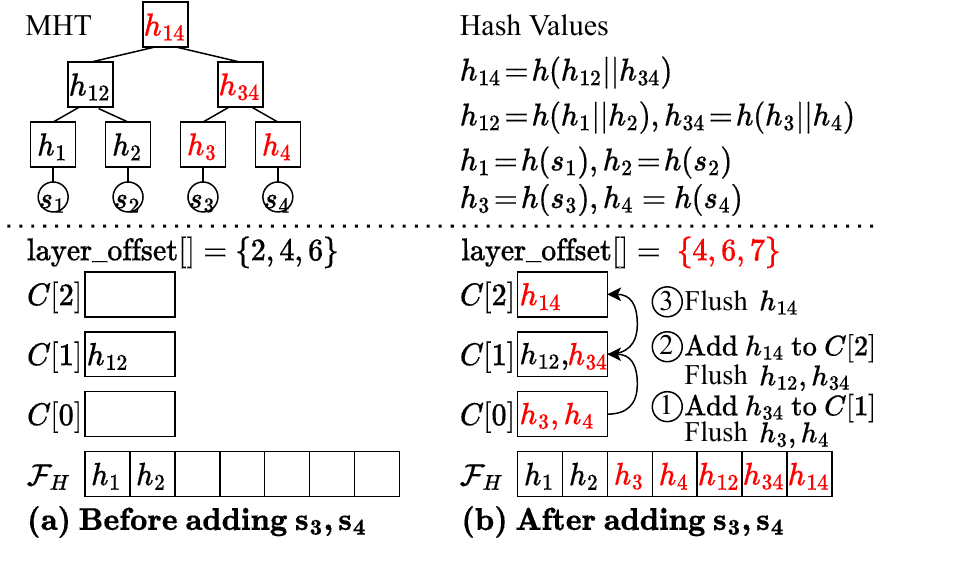}
		\vspace{-0.5em}
		\caption{An Example of Merkle File Construction}\label{fig:merkle-file-example}
	\end{figure}
	
	\begin{example}
		\Cref{fig:merkle-file-example} shows an example of a 2-ary MHT with states $s_1$ to $s_4$. \ce{According to the MHT's structure, $N_{nodes} = [4, 2, 1]$ and $\text{layer\_offset} = [0, 4, 6]$. Assume that $s_1, s_2$ are already added. In this case, $\mathcal{F}_H$ has $h_1, h_2$ and cache $C[1]$ contains $h_{12}$, where $h_1, h_2$ are the hash values of $s_1, s_2$ and $h_{12}$ is derived from $h_1, h_2$ (\cref{fig:merkle-file-example}(a)). Meanwhile, $\text{layer\_offset}[0]$ has been updated to $2$.} After $s_3, s_4$ are added, their hashes $h_3, h_4$ will be inserted to cache $C[0]$, resulting in $C[0]$ having $2$ hash values. Thus, $h_{34}$ derived from $h_3, h_4$ will be added into cache $C[1]$ and $h_3, h_4$ are then flushed to $\mathcal{F}_H$ at offset $2$ (step \raisebox{.5pt}{\textcircled{\raisebox{-.9pt} {1}}}). Since $C[1]$ also has $2$ hash values so the derived $h_{14}$ is added to cache $C[2]$ and $h_{12}, h_{34}$ are flushed to $\mathcal{F}_H$ at offset $\text{layer\_offset}[1] = 4$ (step \raisebox{.5pt}{\textcircled{\raisebox{-.9pt} {2}}}). Finally, $h_{14}$ in $C[2]$ is flushed to $\mathcal{F}_H$ at offset $\text{layer\_offset}[2] = 6$  (step \raisebox{.5pt}{\textcircled{\raisebox{-.9pt} {3}}}).
	\end{example}
	
	\ce{\subsection{Discussions}
		As discussed earlier, COLE adopts the LSM-tree-based maintenance approach to optimize data writes and disk operations under the column-based design. However, it also comes with some tradeoffs. The presence of multiple levels can impact read performance, as retrieving a state requires traversing multiple levels until a satisfactory result is found. Additionally, the merge operation complicates the process of state rewind, as data cannot be deleted in-place. Therefore, COLE does not support blockchain forking and is designed to work with blockchains that do not fork~\cite{gilad2017algorand, wood2016polkadot, androulaki2018hyperledger}.
		
		We next discuss the ACID properties in COLE. COLE achieves atomicity by maintaining \textsf{root\_hash\_list} in an atomic manner. During the level merge process, \textsf{root\_hash\_list} is updated atomically only after constructing all three files in the new level, followed by removing the old level files. This ensures data consistency as the old level files remain intact and are referenced by \textsf{root\_hash\_list} even during a node crash. Concurrency control is not required due to the write-serializability guarantee of the consensus protocol. Data integrity is ensured using Merkle-based structures for each level. For durability, COLE uses transaction logs as the Write Ahead Log since they are agreed upon by the consensus protocol. In case of a crash, COLE recovers by replaying transactions since the last checkpoint. A checkpoint is created when the in-memory MB-tree is flushed to the first disk level and cleared. At this time point, all the data in the system is safely stored on the disk. After a crash, COLE reverts to the last checkpoint, discards all the files in the unfinished merge levels, and starts fresh with an empty in-memory MB-tree. It then replays all unprocessed transactions and restarts the aborted level merges.

	}

\section{Write with Asynchronous Merge}\label{sec:async-write}
\begin{figure*}[t]
	\centering
	\begin{minipage}[t]{.275\textwidth}
		\includegraphics[width=\linewidth]{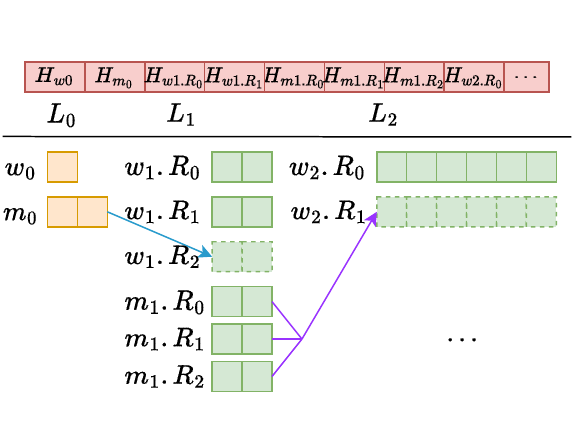}
		\vspace{-1.5em}
		\caption{Asynchronous Merge}\label{fig:COLE-structure-async}
	\end{minipage}
	\begin{minipage}[t]{.695\textwidth}
		\includegraphics[width=\linewidth]{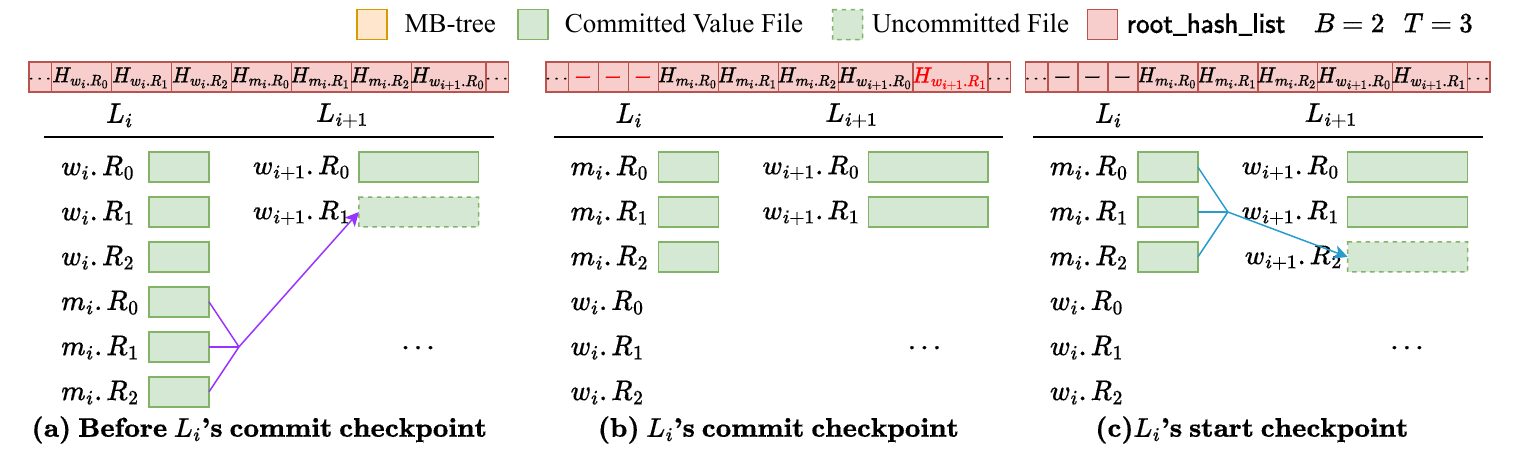}
		\vspace{-1.5em}
		\caption{An Example of Asynchronous Merge}\label{fig:async-new-example}
	\end{minipage}
\end{figure*}

\Cref{alg:complete-insertion} may trigger recursive merge operations during some writes (e.g., steps~\raisebox{.5pt}{\textcircled{\raisebox{-.9pt} {1}}} and \raisebox{.5pt}{\textcircled{\raisebox{-.9pt} {2}}} in \cref{fig:level-merge}). \ce{As a result, it can introduce long-tail latency and cause all future operations to stall. This issue is known as \emph{write stall}, which leads to periodic drops in application throughput to near-zero levels and dramatic fluctuations in system performance.} A common solution is to make the merge operations asynchronous by moving them to separate threads. However, the existing asynchronous merge solution is not suitable for blockchain applications. Since different nodes in the blockchain network could have drastically different computation capabilities, the storage structure will become out-of-sync among nodes when applying asynchronous merges. This will result in different $H_{state}$'s and break the requirement of the blockchain protocol.

\ce{To address these challenges, we design a novel asynchronous merge algorithm for COLE, which ensures the synchronization of the storage across blockchain nodes.} The algorithm introduces two checkpoints, \emph{start} and \emph{commit}, within the asynchronous merge process for each on-disk level. By synchronizing the checkpoints, we ensure consistent blockchain storage and thus $H_{state}$ agreed by the network. \ce{To further minimize the possibility of long-tail latency due to delays at the commit checkpoint, we propose to make the interval between the start checkpoint and the commit checkpoint proportional to the size of the run. This ensures that the majority of the nodes in the network can complete the merge operation before reaching the commit checkpoint.}

\begin{algorithm}[t]
	\caption{Write Algorithm with Asynchronous Merge}\label{alg:async-insertion}
	\SetKwFunction{FMain}{Put}
	\SetKwBlock{thread}{start thread do}{end}
	\Fn{\FMain{$addr, value$}}{
		\KwIn{State address $addr$, value $value$}
		$blk\gets$ current block height;
		$\mathcal{K}\gets \langle addr, blk\rangle$\label{alg:async:1}\;
		$w_{0}\gets$ Get $L_0$'s writing group\;
		Insert $\langle \mathcal{K}, value \rangle$ into the MB-tree of $w_0$\label{alg:async:2}\;
		$i \gets 0$\;
		\While{$w_i$ becomes full}{
			$m_i \gets$ Get $L_i$'s merging group\;
			\If{$m_i.\text{merge\_thread}$ exists}{
				Wait for $m_i.\text{merge\_thread}$ to finish\label{alg:async:wait}\;
				Add the root hash of the generated run from $m_i.\text{merge\_thread}$ to \textsf{root\_hash\_list}\label{alg:async:commit-1}\;
				Remove the root hashes of the runs in $m_i$ from \textsf{root\_hash\_list}\label{alg:async:commit-2}\;
				Remove all the runs in $m_i$\label{alg:async:commit-3}\;
			}
			Switch $m_i$ and $w_i$\label{alg:async:switch}\;
			$m_i.\text{merge\_thread} \gets$~\label{alg:async:start-1}\thread{
				\eIf{$i = 0$}{
					Flush the leaf nodes in $m_i$ to $L_{i+1}$'s writing group a sorted run\;
					Generate files $\mathcal{F}_{V}, \mathcal{F}_{I}, \mathcal{F}_{H}$ for the new run\;
				}{
					Sort-merge all the runs in $m_i$ to $L_{i+1}$'s writing group a new run\;
					Generate files $\mathcal{F}_{V}, \mathcal{F}_{I}, \mathcal{F}_{H}$ for the new run\;
				}\label{alg:async:start-2}
			}
			$i \gets i+1$\;
		}
		Update $H_{state}$ when finalizing the current block\label{alg:async:hash}\;
	}
\end{algorithm}
To realize our idea, we propose to have each level of COLE contain two groups of runs as shown in \cref{fig:COLE-structure-async}. Each group's design is identical to the one discussed in \cref{sec:write-operation}. Specifically, the in-memory level now contains two groups of MB-tree, each with a capacity of $B$ states. Similarly, each on-disk level contains two groups of up to $T$ sorted runs. \ce{Level $i$ can hold a maximum of $2 \cdot B\cdot T^i$ states.} The two groups in each level have two mutually exclusive roles, namely \emph{writing} and \emph{merging}. The writing group accepts newly created runs from the upper level. On the other hand, the merging group generates a new run from its own data and adds to the writing group of the next level in an asynchronous fashion.

\Cref{alg:async-insertion} shows the write operation in COLE with asynchronous merge. First, new state values are inserted into the current writing group of in-memory level $L_0$ (\crefrange{alg:async:1}{alg:async:2}). \ce{The levels in COLE are then traversed from smaller to larger. When a level is full, we commit the previous merge operation in the current level and start a new merge operation in a new thread. To accommodate slow nodes in the network, we check if the previous merging thread of the current level exists and is still in progress, and wait for it to finish if necessary (\cref{alg:async:wait}).} The previous merge operation is committed by adding the root hash of the newly generated run to \textsf{root\_hash\_list} (\cref{alg:async:commit-1}), while obsolete run hashes are removed from \textsf{root\_hash\_list} (\cref{alg:async:commit-2}) and the obsolete runs in the merging group are also removed (\cref{alg:async:commit-3}). \ce{The above procedure ensures the commit checkpoint occurs simultaneously across nodes in the network, which is essential to synchronize the blockchain states and the corresponding root digest. Following this, the roles between the two groups in the current level are switched (\cref{alg:async:switch}). This means that future write operations will be directed to the vacated space of the new writing group, whereas the merge operation will be performed on the new merging group, which is now full.} The latter starts a new merge thread, whose procedure is similar to that of \cref{alg:complete-insertion} (\crefrange{alg:async:start-1}{alg:async:start-2}). Lastly, when finalizing the current block, $H_{state}$ is updated using stored root hashes in \textsf{root\_hash\_list} (\cref{alg:async:hash}).

\begin{example}
	\Cref{fig:async-new-example} shows an example of the asynchronous merge from level $L_i$ to $L_{i+1}$, where $T=3$. The uncommitted files are denoted by dashed boxes. \Cref{fig:async-new-example}(a) shows COLE's structure before $L_i$'s commit checkpoint, when $L_i$'s writing group $w_i$ becomes full. In case $m_i$'s merging thread (denoted by the purple arrow) is not yet finished, we wait for it to finish. Then, during $L_i$'s commit checkpoint, $w_{i+1}.R_1$'s root hash is added to \textsf{root\_hash\_list} and all runs in $m_i$ (i.e., $m_i.R_0, m_i.R_1, m_i.R_2$) are removed (\cref{fig:async-new-example}(b)). Next, $m_i$ and $w_i$'s roles are switched. Finally, a new thread will be started (denoted by the blue arrow) to merge all runs in $m_i$ to $L_{i+1}$'s writing group as the third run $w_{i+1}.R_2$ (\cref{fig:async-new-example}(c)).
\end{example}

\textbf{Soundness Analysis}. 
Next, we show our proposed asynchronous merge operation is sound. Specifically, the following two requirements are satisfied.
\begin{itemize}
	\item The blockchain states' root digest $H_{state}$ is always synchronized among nodes in the blockchain network regardless of how long the underlying merge operation takes.
	\item The interval between the start checkpoint and the commit checkpoint for each level is proportional to the size of the runs to be merged.
\end{itemize}
The first requirement ensures blockchain states are solely determined by the current committed states and are independent of individual node performance variations. The second requirement minimizes the likelihood of nodes waiting for merge operations of longer runs. We now prove that our algorithm complies with the requirements.

\begin{proof}[Proof Sketch]
	It is trivial to show that the first requirement is satisfied as the update of \textsf{root\_hash\_list} (hence $H_{state}$) occurs outside the asynchronous merge thread, making the update of $H_{state}$ fully synchronous and deterministic. For the second requirement, the interval between the start checkpoint and the commit checkpoint in any level equals the time taken to fill up the writing group in the same level. Since the latter contains those runs to be merged in this level, the interval is proportional to the size of the runs.
\end{proof}

\section{Read Operations of COLE}\label{sec:read-operation}
In this section, we discuss the read operations of COLE, including the get query and the provenance query with its verification function. We assume that COLE is implemented with the asynchronous merge.
\subsection{Get Query}
\begin{algorithm}[t]
	\caption{Get Query}\label{alg:simple-lookup}
	\SetKwFunction{FMain}{Get}
	\Fn{\FMain{$addr$}}{
		\KwIn{State address $addr$}
		\KwOut{State latest value $value$}
		$\mathcal{K}_q\gets \langle addr, max\_int \rangle$\;\label{alg:simple-lookup-1}
		
		\ForEach{$g$~\textbf{in}~$\{L_0$'s writing group, $L_0$'s merging group$\}$}{\label{alg:simple-lookup-2}
			$\langle K', state' \rangle \gets \FuncSty{SearchMBTree}(g, \mathcal{K}_q)$\;
			\lIf{$\mathcal{K}'.addr = addr$}{\Return{$state'$}}\label{alg:simple-lookup-3}
		}
		\ForEach{level $i$ \textbf{in} $\{1, 2, \dotsc\}$}{\label{alg:simple-lookup-4}
			$RS\!\gets\!\{R_{i,j}~|~R_{i,j}\!\in\!L_i\text{'s writing group} \land \text{committed}\}$\;
			$RS\!\gets\!RS + \{ R_{i,j}~|~R_{i,j}\!\in\!L_i$'s merging group$ \}$\;
			
			\ForEach{$R_{i,j}$ \textbf{in} $RS$}{
				$\langle \langle K', state'\rangle, pos' \rangle \gets \FuncSty{SearchRun}(R_{i,j}, \mathcal{K}_q)$\;
				\lIf{$\mathcal{K}'.addr = addr$}{\Return{$state'$}}\label{alg:simple-lookup-5}
			}
		}
		\Return{$nil$}\;
	}
\end{algorithm}

\begin{algorithm}[t]
	\caption{Search a Run}\label{alg:search-run}
	\SetKwFunction{FMain}{SearchRun}
	\SetKwFunction{FSub}{QueryModel}
	\Fn{\FMain{$\mathcal{F}_{I}, \mathcal{F}_{V}, \mathcal{B}, \mathcal{K}_q$}}{
		\KwIn{Index file $\mathcal{F}_{I}$, value file $\mathcal{F}_{V}$, bloom filter $\mathcal{B}$, compound key $\mathcal{K}_q = \langle addr_q, blk_q\rangle$}
		\KwOut{Queried state $s$ and its position $pos$}
		\lIf{$addr_q \notin \mathcal{B}$}{\label{alg:search-run-1}\Return{}}
		$\mathcal{K}_q\gets BigNum(\mathcal{K}_q)$\;
		$\mathcal{P}\gets \mathcal{F}_{I}$'s last page;
		$\mathcal{M}\gets \FuncSty{BinarySearch}(\mathcal{P}, \mathcal{K}_q)$\; \label{alg:search-run-2}
		$\langle \mathcal{M}, pos \rangle \gets \FuncSty{QueryModel}(\mathcal{M}, \mathcal{F}_I, \mathcal{K}_q)$\; \label{alg:search-run-3.1}
		\While{$pos$ is \textbf{not} pointing to the bottom models}{
			$\langle \mathcal{M}, pos \rangle \gets \FuncSty{QueryModel}(\mathcal{M}, \mathcal{F}_I, \mathcal{K}_q)$\;\label{alg:search-run-3.2}
		}
		\Return{$\FuncSty{QueryModel}(\mathcal{M}, \mathcal{F}_V, \mathcal{K}_q)$}\;\label{alg:search-run-3.3}
	}
	
	\Fn{\FSub{$\mathcal{M}, \mathcal{F}, \mathcal{K}_q$}}{
		\KwIn{Model $\mathcal{M}$, query file $\mathcal{F}$, compound key $\mathcal{K}_q$}
		\KwOut{Queried data and its position in $\mathcal{F}$}
		$\langle sl, ic, k_{min}, p_{max} \rangle \gets \mathcal{M}$\;
		\lIf{$\mathcal{K}_q < k_{min}$}{\label{alg:search-run-4}\Return{}}
		$pos_{pred} \gets \min(\mathcal{K}_q \cdot sl + ic, p_{max})$\;\label{alg:search-run-5}
		$page_{pred} \gets pos_{pred}/2\epsilon$\;\label{alg:search-run-6}
		$\mathcal{P}\gets \mathcal{F}\text{'s page at } page_{pred}$\;
		\uIf{$\mathcal{K}_q < \mathcal{P}[0].k$}{\label{alg:search-run-7}
			$\mathcal{P}\gets \mathcal{F}\text{'s page at } page_{pred} - 1$\;
		}
		\ElseIf{$\mathcal{K}_q > \mathcal{P}[-1].k$}{
			$\mathcal{P}\gets \mathcal{F}\text{'s page at } page_{pred} + 1$\;\label{alg:search-run-8}
		}
		\Return{$\FuncSty{BinarySearch}(\mathcal{P}, \mathcal{K}_q)$}\;\label{alg:search-run-9}
	}
\end{algorithm}

\Cref{alg:simple-lookup} shows the get query process. As mentioned in \cref{sec:design-overview}, getting a state's latest value requires a special compound key $\mathcal{K}_q = \langle addr_q, max\_int\rangle$. \ce{Owing to the temporal order of COLE's levels, we perform the search from smaller levels to larger levels, until a satisfied state value is found.} This involves searching both the writing and merging groups' MB-trees in the in-memory level $L_0$ as both of them are committed (\crefrange{alg:simple-lookup-2}{alg:simple-lookup-3}). Then, in each on-disk level, a search is performed in the committed writing group's runs, followed by the merging group's runs (\crefrange{alg:simple-lookup-4}{alg:simple-lookup-5}). \ce{Note that the runs in the same group will be searched in the order of their freshness.} For the example in \cref{fig:COLE-structure-async}, we search the MB-trees in $w_0$ and $m_0$, followed by the runs in the order of $w_1.R_1$, $w_1.R_0$, $m_1.R_2$, $m_1.R_1$, $m_1.R_0$, $w_2.R_0$, $\cdots$, while the uncommitted $w_1.R_2, w_2.R_1$ are skipped. The search halts once the satisfied state is found.

To search an on-disk run, we use \cref{alg:search-run}. First, if the queried address $addr_q$ is not in the run's bloom filter $\mathcal{B}$, the run is skipped (\cref{alg:search-run-1}). Otherwise, models in the index file $\mathcal{F}_I$ are used to find $\mathcal{K}_q$. The search starts from the top layer of models, stored on the last page of $\mathcal{F}_I$. The model covering $\mathcal{K}_q$ is found by binary searching $k_{min}$ of each model in this page (\cref{alg:search-run-2}). Then, a recursive query on models in subsequent layers is conducted from top to bottom (\crefrange{alg:search-run-3.1}{alg:search-run-3.2}). Upon reaching the bottom layer, the corresponding model is used to locate the state value in the value file $\mathcal{F}_V$ (\cref{alg:search-run-3.3}).

Function $\FuncSty{QueryModel}(\cdot)$ in \cref{alg:search-run} shows the procedure of using a learned model $\mathcal{M}$ to locate the queried compound key $\mathcal{K}_q$. If the model covers $\mathcal{K}_q$, it predicts the position $pos_{pred}$ of the queried data (\cref{alg:search-run-5}). With the error bound of the model $2\epsilon$ equaling the page size, the predicted page id is computed as $pos_{pred}/2\epsilon$ (\cref{alg:search-run-6}). The corresponding page $\mathcal{P}$ is fetched and the first and last models are checked whether they cover $\mathcal{K}_q$. If not, the adjacent page is fetched as $\mathcal{P}$ (\crefrange{alg:search-run-7}{alg:search-run-8}). This process involves at most two pages for prediction, hence minimizing IO. Finally, a binary search in $\mathcal{P}$ locates the queried data (\cref{alg:search-run-9}).

\subsection{Provenance Query}

A provenance query resembles a get query but with notable distinctions. Unlike a get query, a provenance query involves a range search based on the queried block height range. This entails computing two boundary compound keys, $\mathcal{K}_l=\langle addr, blk_l - 1 \rangle$ and $\mathcal{K}_u=\langle addr, blk_u + 1 \rangle$, with offsets adjusted by one to prevent the omission of valid results. Moreover, a provenance query provides Merkle proofs to authenticate the results. 

Specifically, during the search of MB-trees in $L_0$, in addition to retrieving satisfactory results, Merkle paths are included in the proof using a similar approach mentioned in \cref{sec:prelim-blockchain-index}. For the runs of the on-disk levels, we search in the same order as those described in \cref{alg:simple-lookup}. $\mathcal{K}_l$ is used as the search key when applying the learned models to find the first query result in each run. Then, the value file is scanned sequentially until a state beyond $\mathcal{K}_u$ is reached.\footnote{For simplicity, we assume that $addr$ is in the bloom filter $\mathcal{B}$. If not, $\mathcal{B}$ is also added as the proof to prove that $addr$ is not in the run.} Afterwards, a Merkle proof is computed upon the first and last results' positions $pos_l, pos_u$ of each run. Since the states in the value file and their hash values in the Merkle file share the same position, the Merkle paths of the hash values at $pos_l$ and $pos_u$ are used as the Merkle proof. To compute the Merkle path, we traverse the MHT in the Merkle file from bottom to top. Note that given a hash value's position $pos$ at layer $i$, we can directly compute its parent hash value's position in the Merkle file as $\floor{{(pos -  \sum_{0}^{i-1} \ceil{\frac{n}{m^i}})}/{m}} +  \sum_{0}^{i} \ceil{\frac{n}{m^i}}$. Due to the space limitation, the detailed procedure of the provenance query is given in
\cref{sec:appendix}.

On the user's side, the verification algorithm works as follows: (1) use each MB-tree's results and their corresponding Merkle proof to reconstruct the MB-tree's root hash; (2) use each searched run's results and their corresponding Merkle proof to reconstruct the run's root hash; (3) use the reconstructed root hashes to reconstruct the states' root digest and compare it with the published one, $H_{state}$, in the block header; (4) check the boundary results of each searched run against the compound key range $[\mathcal{K}_l, \mathcal{K}_u]$ to ensure no missing results. If all these checks pass, the results are verified.

\section{Complexity Analysis}\label{sec:analysis}

\begin{table}[t]
	\centering
	\resizebox{\linewidth}{!}{
		\small
		\begin{tabular}{|c|c|c|c|}
			\hline
			\textbf{Cost} & \textbf{MPT} & \textbf{COLE}  & \textbf{COLE w/ async-merge} \\
			\hline \hline
			Storage size & $O( n\cdot d_{MPT})$ & \multicolumn{2}{c|}{$O( n )$} \\  \hline
			Write IO cost & $O(d_{MPT})$  &  \multicolumn{2}{c|}{$O(d_{COLE})$} \\ \hline
			Write tail latency & $O(1)$ & $O(n)$ & $O(1)$ \\ \hline
			Write memory footprint & $O(1)$  & $O(T + m\cdot d_{COLE} )$  & $O( T\cdot d_{COLE} + m\cdot d_{COLE}^2 )$ \\ \hline
			Get query IO cost & $O( d_{MPT})$ &  \multicolumn{2}{c|}{$O( T\cdot d_{COLE} \cdot C_{model})$} \\ \hline
			Prov-query IO cost & $O( d_{MPT})$ &  \multicolumn{2}{c|}{$O( T\cdot d_{COLE} \cdot C_{model} + m\cdot d_{COLE}^2)$} \\ \hline
			Prov-query proof size & $O( d_{MPT})$  &  \multicolumn{2}{c|}{$O( m\cdot d_{COLE}^2)$} \\
			\hline
	\end{tabular}}
	\vspace{-0.5em}
	\caption{Complexity Comparison}\label{tab:complexity}
\end{table}

In this section, we analyze the complexity in terms of storage, memory footprint, and IO cost. To ease the analysis, we assume $n$ as the total historical values, $T$ as the level size ratio, $B$ as the in-memory level's capacity, and $m$ as COLE's MHT fanout. \Cref{tab:complexity} shows the comparison of MPT, COLE, and COLE with the asynchronous merge.


\ce{We first analyze the storage size. Since MPT duplicates the nodes of the update path for each insertion, its storage has a size of $O(n\cdot d_{MPT})$, where $d_{MPT}$ is the height of the MPT. COLE completely removes the node duplication, thus achieving an $O(n)$ storage size.
	
	Next, we analyze the write IO cost. MPT takes $O(d_{MPT})$ to write the nodes in the update path, while COLE takes $O(d_{COLE})$ for the worst case when all levels are merged, where $d_{COLE}$ is the number of levels in COLE. Similar to the traditional LSM-tree's write cost~\cite{dayan2018dostoevsky}, the level merge in COLE takes an amortized $O(1)$ IO cost to write the value file, the index file, and the Merkle file. The number of levels $d_{COLE}$ is $\ceil{\log_T (\frac{n}{B}\cdot \frac{T-1}{T})}$, which is logarithmic to $n$.} Note that normally $d_{COLE}<d_{MPT}$ since $d_{MPT}$ depends on the data's key size, which can be large (e.g., when having a 256-bit key, maximum $d_{MPT}$ is 64 under hexadecimal base while COLE has only a few levels following the LSM-tree).

Regarding the write tail latency, MPT has a constant cost since there is no write stall during data writes. On the other hand, COLE may experience the write stall in the worst case, which requires waiting for the merge of all levels and results in the reading and writing of $O(n)$ states. The asynchronous merge algorithm removes the write stall by merging the levels in background threads and reduces the tail latency to $O(1)$.

As for the write memory footprint, MPT has a constant cost since the update nodes are computed on the fly and can be removed from the memory after being flushed to the disk. For COLE, we consider the case of merging the largest level as this is the worst case. The sort-merge takes $O(T)$ memory and the model construction takes constant memory~\cite{o1981line}. Constructing the Merkle file takes $O(m\cdot d_{COLE})$ since there are logarithmic layers of cache buffers and each buffer contains $m$ hash values. To sum up, COLE takes $O(T + m\cdot d_{COLE})$ memory during a write operation. For COLE with the asynchronous merge, the worst case is that each level has a merging thread, thus requiring $d_{COLE}$ times of memory compared with the synchronous merge, i.e., $O( T\cdot d_{COLE} + m\cdot d_{COLE}^2 )$.

We finally analyze the read operations' costs, including the get query IO cost, the provenance query IO cost, and the proof size of the provenance query. MPT's costs are all linear to the MPT's height, $O(d_{MPT})$. For COLE, $T$ runs in each level should be queried, where we assume that each run takes $C_{model}$ to locate the state. Therefore, the cost of the get query is $O( T\cdot d_{COLE} \cdot C_{model})$. To generate the Merkle proof during the provenance query, an additional $O(m\cdot d_{COLE}^2)$ is required since there are multiple layers of MHT in all levels and $O(m)$ hash values are retrieved for each MHT's layer. The proof size is $O(m\cdot d_{COLE}^2)$ for a similar reason.

\section{Evaluation}\label{sec:evaluation}
In this section, we first describe the experiment setup, including comparing baselines, implementation, parameter settings, workloads, and evaluation metrics. Then, we present the experiment results.

\subsection{Experiment Setup}

\subsubsection{Baselines}

We compare COLE with the following baselines:
\begin{itemize}
	\item \textsf{MPT}: It is used by Ethereum to index the blockchain storage. The structure is made persistent as mentioned in \cref{sec:intro}.
	\item \textsf{LIPP}: It applies LIPP~\cite{wu2021updatable}, the state-of-the-art learned index supporting \emph{in-place} data writes, to the blockchain storage without our column-based design. \textsf{LIPP} retains the node persistence strategy to support provenance queries.
	\item \textsf{Column-based Merkle Index (CMI)}: It uses the column-based design with traditional Merkle indexes rather than the learned index. It adopts a two-level structure. The upper index is a non-persistent MPT whose key is the state address and the value is the root hash of the lower index. The lower index follows the column-based design, using an MB-tree to store the state's historical values in a contiguous fashion~\cite{li2006dynamic}.
\end{itemize}

\subsubsection{Implementation and Parameter Setting}

We implement COLE and the baselines in Rust programming language. The source code is available at~\url{https://github.com/hkbudb/cole}. We use the Rust Ethereum Virtual Machine (EVM) to execute transactions, simulating blockchain data updates and reads~\cite{evm}. Transactions are packed into blocks, each containing 100 transactions. Ten smart contracts are initially deployed and repeatedly invoked with transactions. Big number operations mentioned in \cref{sec:design-overview} are implemented using the \emph{rug} library~\cite{rug}. Baselines utilize RocksDB~\cite{rocksdb} as the underlying storage, while COLE uses simple files for data storage as enabled by our design.

We set $\epsilon = 23$ based on the page size (4KB) and the compound key-pair size (88 bytes). By default, the size ratio $T$ and the MHT fanout $m$ of COLE are set to 4. \ce{Following the default configuration of RocksDB, its memory budget is set to 64MB.} The in-memory capacity $B$ is set to the number of states that can fit within the same memory budget. \Cref{tab:params} shows all the parameters where the default settings are highlighted in bold font. All experiments are run on a machine equipped with an Intel i7-10710U CPU, 16GB RAM, and Samsung SSD 256GB.

\begin{table}[t]
	\small
	\centering
	\begin{tabular}{ll}
		\toprule
		\textbf{Parameters}    & \textbf{Value}               \\ \midrule
		\# of generated blocks & $10^2, 10^3, 10^4, \mathbf{10^5}$ \\
		Size ratio $T$ & $2, \mathbf{4}, 6, 8, 10, 12$ \\
		COLE's MHT fanout $m$     & $2, \mathbf{4}, 8, 16, 32, 64$           \\
		\bottomrule
	\end{tabular}
	\caption{System Parameters}%
	\label{tab:params}
\end{table}

\subsubsection{Workloads and Evaluation Metrics}\label{sec:workloads-metrics}

The experiment evaluation includes two parts: the overall performance of transaction executions and the performance of provenance queries. For the first part, SmallBank and KVStore from Blockbench~\cite{dinh2017blockbench} are used as macro benchmarks to generate the transaction workload. SmallBank simulates the account transfers while KVStore uses YCSB~\cite{cooper2010benchmarking} for read/write tests. YCSB involves a loading phase where base data is generated and stored, followed by a running phase for read/update operations. A transaction that reads/updates data is denoted as a read/write transaction. We set $10^5$ transactions as the base data and vary read/update ratios to simulate different scenarios: (i) Read-Write with equal read/write transactions; (ii) Read-Only with only read transactions; and (iii) Write-Only with all write transactions. The overall performance is evaluated in terms of the average transaction throughput, the tail latency, and the storage size.

To evaluate provenance queries, we use KVStore to simulate the workload including frequent data updates. We initially write 100 states as the base data and then continuously generate write transactions to update the base data's states. For each query, we randomly select a key from the base data and vary the block height range (e.g., $2, 4, \cdots, 128$), which follows \cite{ruan2019fine}'s setting. The evaluation metrics include (i) CPU time of the query executed on the blockchain node and verified by the query user and (ii) proof size.

\subsection{Experimental Results}

\subsubsection{Overall Performance}

\begin{figure}[t]
	\centering
	\includegraphics[width=.49\linewidth]{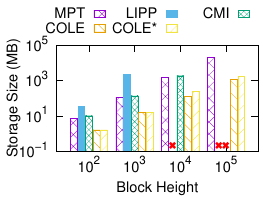}~%
	\includegraphics[width=.49\linewidth]{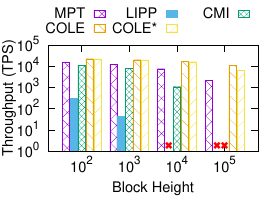}
	\vspace{-1em}
	\caption{Performance vs. Block Height (SmallBank)}%
	\label{fig:smallbank-perf}
\end{figure}

\begin{figure}[t]
	\centering
	\includegraphics[width=.49\linewidth]{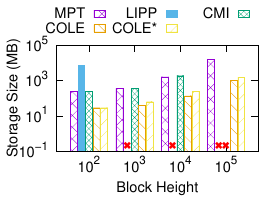}~%
	\includegraphics[width=.49\linewidth]{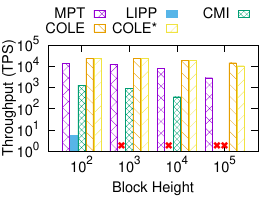}
	\vspace{-1em}
	\caption{Performance vs. Block Height (KVStore)}%
	\label{fig:kvstore-perf}
\end{figure}

\Cref{fig:smallbank-perf,fig:kvstore-perf} show the storage size and throughput of COLE and all baselines under the SmallBank and KVStore workloads, respectively. We denote COLE with the asynchronous merge as COLE*.

We make several interesting observations. 
First, COLE significantly reduces the storage size compared to MPT as the blockchain grows. For example, at a block height of $10^5$, the storage size decreases by 94\% and 93\% for SmallBank and KVStore, respectively. This is due to COLE's elimination of the need to persist internal data structures via the column-based design, and its use of storage-efficient learned models for indexing. Moreover, COLE outperforms MPT in throughput, achieving a $1.4\times$-$5.4\times$ improvement, thanks to its learned index. COLE* performs slightly worse than COLE due to the overhead of the asynchronous merge. 

Second, using the learned index without the column-based design (\textsf{LIPP}) even increases the blockchain storage. At a block height of $10^2$, the storage size of \textsf{LIPP} already exceeds \textsf{MPT}'s by $5\times$ (for SmallBank) and $31\times$ (for KVStore). This happens because the learned index often generates larger index nodes that must be persisted with each new block, leading to increased storage and significant IO operations. Consequently, \textsf{LIPP}'s throughput is significantly worse than \textsf{MPT}. We are not able to report the results of \textsf{LIPP} for the block height above $10^3$ for SmallBank and $10^2$ for KVStore as the experiment could not be finished within 24 hours.

Third, extending \textsf{MPT} with the column-based design (\textsf{CMI}) does not significantly change the storage size. The additional storage of the lower-level MB-tree and the use of the RocksDB backend largely negate the benefit of removing node persistence.  Additionally, refreshing Merkle hashes of all nodes in the index update path, which entails both read and write IOs, further impacts performance. Consequently, the throughput of \textsf{CMI} is $7\times$ and $22\times$ worse than \textsf{MPT} for SmallBank and KVStore, respectively, at a block height of $10^4$. The experiments of \textsf{CMI} cannot scale beyond a block height of $10^4$.

Overall, with a unique combination of the learned index, column-based design, and write-optimized strategies, COLE and COLE* not only achieve the smallest storage requirement but also gain the highest system throughput.

\begin{figure}[t]
	\includegraphics[width=.49\linewidth]{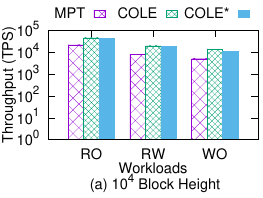}
	\includegraphics[width=.49\linewidth]{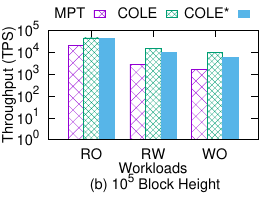}
	\vspace{-1em}
	\caption{Throughput vs. Workloads (KVStore)}\label{fig:workloads-throughput}
\end{figure}

\subsubsection{Impact of Workloads}
We use KVStore to evaluate the impact of different workloads, namely Read-Only (RO), Read-Write (RW), and Write-Only (OW), in terms of the system throughput. As shown in \cref{fig:workloads-throughput}, the throughputs of all systems decrease with more write operations in the workload. The performance of \textsf{MPT} degrades by up to $93\%$ while that of COLE and COLE* degrades by up to $87\%$. This shows that the LSM-tree-based maintenance approach helps optimize the write operation. We omit \textsf{LIPP} and \textsf{CMI} in \cref{fig:workloads-throughput} since they cannot scale beyond a block height of $10^3$ and $10^4$, respectively.

\subsubsection{Tail Latency}
\begin{figure}[t]
	\centering
	\includegraphics[width=.49\linewidth]{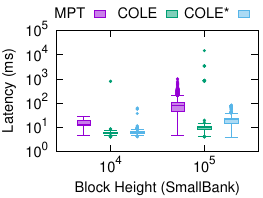}~%
	\includegraphics[width=.49\linewidth]{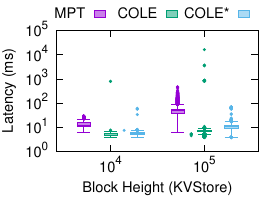}
	\vspace{-1em}
	\caption{Latency Box Plot}%
	\label{fig:tail-latency}
\end{figure}

To assess the effect of the asynchronous merge, \cref{fig:tail-latency} shows the box plot of the latency of SmallBank and KVStore workloads at block heights of $10^4$ and $10^5$. The tail latency is depicted as the maximum outlier. As the blockchain grows, COLE* decreases the tail latency by 1-2 orders of magnitude for both workloads. This shows that the asynchronous merge strategy will become more effective when the system scales up for real-world applications. Owing to the asynchronous merge overhead, COLE* incurs slightly higher median latency than COLE, but it still outperforms MPT.

\subsubsection{Impact of Size Ratio}

\begin{figure}[t]
	\centering
	\includegraphics[width=.49\linewidth]{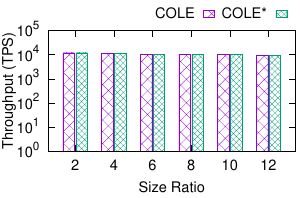}~%
	\includegraphics[width=.49\linewidth]{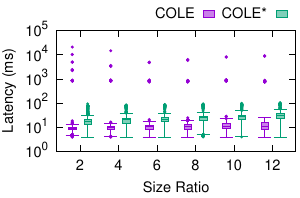}
	\vspace{-1em}
	\caption{Impact of Size Ratio}%
	\label{fig:params-tuning}
\end{figure}

\Cref{fig:params-tuning} shows the system throughput and latency box plot under $10^5$ block height using the SmallBank benchmark with varying size ratio $T$. As the size ratio increases, the throughput remains stable, while the tail latency shows a U shape. We observe that $T=6$ and $T=4$ are the best settings for COLE and COLE*, respectively,  with the lowest tail latency. Meanwhile, with an increasing size ratio, the median latency of both COLE and COLE* increases.

\subsubsection{Provenance Query Performance}

\begin{figure}[t]
	\centering
	\includegraphics[width=.49\linewidth]{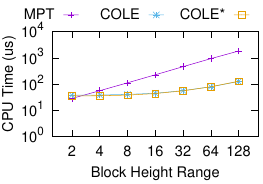}~%
	\includegraphics[width=.49\linewidth]{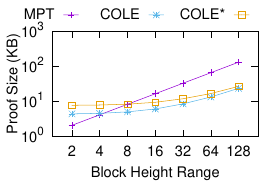}
	\vspace{-1em}
	\caption{Prov-Query Performance vs. Query Range}%
	\label{fig:prov}
\end{figure}


We now evaluate the performance of provenance queries by querying historical state values of a random address within the latest $q$ blocks. With the current block height fixed at $10^5$, we vary $q$ from $2$ to $128$. \textsf{LIPP} and \textsf{CMI} are omitted here since they cannot scale at $10^5$ block height. \Cref{fig:prov} shows that \textsf{MPT}'s CPU time and proof size grow linearly with $q$ while those of COLE and COLE* grow only sublinearly. This is because \textsf{MPT} requires to query each block inside the queried range. In contrast, COLE and COLE*'s column-based design often locates query results within contiguous storage of each run, hence reducing the number of index traversals during the query and shrinking the proof size by sharing ancestor nodes in the Merkle path. COLE and COLE*'s proof sizes surpass that of \textsf{MPT} when the query range is small due to limited sharing capabilities within a small query range.

\section{Related Work}\label{sec:relatedworks}
In this section, we briefly review the related works on learned indexes and blockchain storage management.

\subsection{Learned Index}
Learned index has been extensively studied in recent years.
The original learned index~\cite{kraska2018case} only supports static data while PGM-index~\cite{ferragina2020pgm}, Fiting-tree~\cite{galakatos2019fiting}, ALEX~\cite{ding2020alex}, LIPP~\cite{wu2021updatable}, and LIFOSS~\cite{yu2022lifoss} support dynamic data using different strategies. All these works are designed and optimized for in-memory databases. Bourbon~\cite{dai2020wisckey} uses the PGM-based models to speed up the lookup in the WiscKey system, which is a persistent key-value store. \cite{lan2023updatable} investigates how existing dynamic learned indexes perform on-disk and shows the design choices. Some other learned indexes are proposed for more complex application scenarios like spatial data~\cite{li2020lisa, gu2021reinforcement, sheng2023wisk}, multi-dimensional data~\cite{nathan2020learning, tsunami20}, and variable-length string data~\cite{wang2020sindex}. Moreover, \cite{li2021finedex, lu2021apex} consider designing learned indexes for concurrent systems. \ce{\cite{zhang2022plin} proposes a persistent learned index that is specifically designed for the NVM-only architecture with concurrency control.} More recently, \cite{li2023rolex} designs a scalable RDMA-oridented learned key-value store for disaggregated memory systems. Nevertheless, existing works cannot be directly applied to blockchain storage since they do not take into account disk-optimized storage, data integrity, and provenance queries simultaneously.

\subsection{Blockchain Storage Management}
Pioneering blockchain systems, such as Bitcoin~\cite{nakamoto2008bitcoin} and Ethereum~\cite{wood2014ethereum}, use MPT and store it using simple key-value storage like RocksDB~\cite{rocksdb}, which implements the LSM-tree structure. While many works propose to optimize the generic LSM-tree for high throughput and low latency~\cite{dayan2022spooky, sarkarconstructing, dayan2018dostoevsky, 10.1145/3035918.3064054, yu2023adoc}, and some propose orthogonal designs that could potentially be incorporated into COLE, they are not specifically designed to meet the unique integrity and provenance requirements of blockchain systems. On the other hand, a large body of research has been carried out to study alternative solutions to reduce blockchain storage overhead. Several studies~\cite{zamani2018rapidchain, el2019blockchaindb, dang2019towards, resilientdb20, honggridb23} consider using \emph{sharding} techniques to horizontally partition the blockchain storage and each partition is maintained by a subset of nodes, thus reducing the overall storage overhead.  Distributed data storage~\cite{xu2018cub, qi2020bft} or moving on-chain states to off-chain nodes~\cite{chepurnoy2018edrax, boneh2019batching, tomescu2020aggregatable, xu2021slimchain, xu2022l2chain} has also been proposed to reduce each blockchain node's storage overhead. Besides, ForkBase~\cite{wang2018forkbase} proposes to optimize blockchain storage by deduplicating multi-versioned data and supporting efficient fork operations. \cite{li2023lvmt} employs a vector commitment protocol and multi-level authenticated trees to reduce I/O costs for blockchain storage. To the best of our knowledge, COLE is the first work that targets the index itself to address the blockchain storage overhead.

Another related topic is to support efficient queries in blockchain systems. LineageChain~\cite{ruan2019fine} focuses on provenance queries in the blockchain. 
Verifiable boolean range queries are studied in vChain and vChain+~\cite{xu2019vchain,wang2022vchain+}, where accumulator-based authenticated data structures are designed. GEM$^2$-tree~\cite{zhang2019gem} explores query processing in the context of on-chain/off-chain hybrid storage. FalconDB~\cite{peng2020falcondb} combines the blockchain and the collaborative database to support SQL queries with a strong security guarantee. \cite{xu2023empowering} studies the authenticated spatial and keyword queries in blockchain databases. iQuery~\cite{lu2022iquery} supports intelligent blockchain analytical queries and guarantees the trustworthiness of query results by using multiple service providers. While all these works focus on proposing additional data structures to process specific queries, COLE  focuses on improving the performance of the general blockchain storage system.

\section{Conclusion}\label{sec:conclusion}
In this paper, we have designed COLE, a novel column-based learned storage for blockchain systems. COLE follows the column-based database design to contiguously store each state's historical values using an LSM-tree approach.
Within each run of the LSM-tree, a disk-optimized learned index has been designed to facilitate efficient data retrieval and provenance queries. Moreover, a streaming algorithm has been proposed to construct Merkle files that are used to ensure blockchain data integrity. In addition, a new checkpoint-based asynchronous merge strategy has been proposed to tackle the long-tail latency issue for data writes in COLE. Extensive experiments show that, compared with the existing systems, the proposed COLE system reduces the storage size by up to 94\% and improves the system throughput by $1.4\times$-$5.4\times$. Additionally, the proposed asynchronous merge decreases the long-tail latency by 1-2 orders of magnitude while maintaining a comparable storage size.

For future work, we plan to extend COLE to support blockchain systems that undergo forking, where the states of a forked block can be rewound. We will investigate efficient strategies to remove the rewound states from storage. Furthermore, since the column-based design stores blockchain states contiguously, compression techniques can be applied to take advantage of similarities between adjacent data. We will study how to incorporate compression strategies into the learned index.

\section*{Acknowledgments}
This work is supported by Hong Kong RGC Grants (Project No. 12200022, 12201520, C2004-21GF). Jianliang Xu is the corresponding author.
\newpage
\bibliographystyle{plain}
\bibliography{ref.bib}
\appendix

\section{Appendix}%
\label{sec:appendix}
\subsection{Provenance Query Algorithm}
\begin{algorithm}[t]
	\caption{Provenance Query}\label{alg:simple-authen-prov}
	\SetKwFunction{FMain}{ProvQuery}
	\Fn{\FMain{$addr, [blk_l, blk_u]$}}{
		\KwIn{State address $addr$, block height range $[blk_l, blk_u]$}
		\KwOut{Result set $R$, proof $\pi$}
		$\mathcal{K}_l\gets \langle addr, blk_l - 1 \rangle$;
		$\mathcal{K}_u\gets \langle addr, blk_u + 1 \rangle$\;\label{alg:simple-authen-prov-1}
		
		\ForEach{$g$~\textbf{in}~$\{L_0$'s writing group, $L_0$'s merging group$\}$}{
			$\langle R', \pi' \rangle \gets \FuncSty{SearchMBTree}(g, [\mathcal{K}_l, \mathcal{K}_u])$\;\label{alg:simple-authen-prov-4.1}
			$R.add(R')$; $\pi.add(\pi')$\; \label{alg:simple-authen-prov-4.2}
			\If{$\min(\{r.blk | r \in R'\}) < blk_l$}{\label{alg:simple-authen-prov-6}
				$\pi.add(\text{remaining of \textsf{root\_hash\_list}})$\;
				\Return{$\langle R, \pi \rangle$}\;\label{alg:simple-authen-prov-7}
			}
		}
		
		\ForEach{level $i$ \textbf{in} $\{1, 2, \dotsc\}$}{
			$RS\!\gets\!\{R_{i,j}~|~R_{i,j}\!\in\!L_i\text{'s writing group}\!\land\!R_{i,j}\text{~is committed}\}$\;
			$RS\!\gets\!RS + \{ R_{i,j}~|~R_{i,j}\!\in\!L_i$'s merging group$ \}$\;
			
			\ForEach{$R_{i,j}$ \textbf{in} $RS$}{
				$\langle \langle \mathcal{K}', state' \rangle, pos_l \rangle \gets \FuncSty{SearchRun}(R_{i,j}, \mathcal{K}_l)$\;\label{alg:simple-authen-prov-2}
				$pos_u \gets pos_l$\; \label{alg:simple-authen-prov-3.1}
				\While{$\mathcal{F}_V[pos_u].k \le \mathcal{K}_u$}{
					$R.add(\mathcal{F}_V[pos_u])$\;
					$pos_u \gets pos_u + 1$\; \label{alg:simple-authen-prov-3.2}
				}
				$\pi.add(\text{MHT proof w.r.t.~} pos_l {~to~} pos_u)$\;\label{alg:simple-authen-prov-5}
				\If{$\mathcal{K}'.blk < blk_l$}{\label{alg:simple-authen-prov-8}
					$\pi.add(\text{remaining of \textsf{root\_hash\_list}})$\; \label{alg:simple-authen-prov-10}
					\Return{$\langle R, \pi \rangle$}\;\label{alg:simple-authen-prov-9}
				}
			}
		}
		\Return{$\langle R, \pi \rangle$}\;
	}
\end{algorithm}
\Cref{alg:simple-authen-prov} shows the procedure of the provenance query. First, we compute two boundary compound keys $\mathcal{K}_l=\langle addr, blk_l - 1 \rangle, \mathcal{K}_u=\langle addr, blk_u + 1 \rangle$ (\cref{alg:simple-authen-prov-1}). The offsets by one are needed to ensure that no valid results will be missing. Then, similar to the get query,  we traverse both MB-trees in $L_0$ to find the results in the query range (\cref{alg:simple-authen-prov-4.1}). At the same time, the corresponding MB-tree paths are added to $\pi$ as the Merkle proof (\cref{alg:simple-authen-prov-4.2}). If we find a state whose block height is smaller than $blk_l$, we stop the search since all states in the following levels must be even older than $blk_l$ (\crefrange{alg:simple-authen-prov-6}{alg:simple-authen-prov-7}). Otherwise, we continue to search the on-disk runs in the same order as those described in \cref{alg:simple-lookup}. We use $\mathcal{K}_l$ as the search key when applying the learned models to find the first query result in each run (\cref{alg:simple-authen-prov-2}). Afterwards, we sequentially scan the value file until the state is outside of the query range based on $\mathcal{K}_u$ (\crefrange{alg:simple-authen-prov-3.1}{alg:simple-authen-prov-3.2}). A Merkle proof is computed accordingly based on the position of the first and the last results in the value file of this run. This proof is added to $\pi$ (\cref{alg:simple-authen-prov-5}). Similar to the in-memory level, we apply an early stop when we find a state's block height is smaller than $blk_l$ (\cref{alg:simple-authen-prov-8}). Finally, the root hashes of the unsearched runs in \textsf{root\_hash\_list} are added to $\pi$ (\cref{alg:simple-authen-prov-10}).

\subsubsection{Impact of COLE'S MHT Fanout}
\begin{figure}[t]
	\centering
	\includegraphics[width=.49\linewidth]{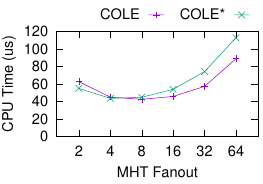}~%
	\includegraphics[width=.49\linewidth]{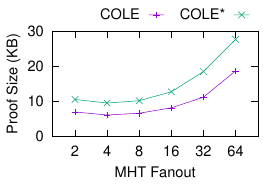}
	\vspace{-1em}
	\caption{Impact of COLE's MHT Fanout}%
	\label{fig:impact-fanout}
\end{figure}
\Cref{fig:impact-fanout} shows the CPU time and proof size under $10^5$ block height and $q=16$ when varying COLE's MHT fanout $m$. We observe a U-shaped trend for both the CPU time and proof size with the increasing fanout. The reason is that as the fanout increases, the MHT height decreases, resulting in shorter CPU time and smaller proof size. However, the size of each node of MHT increases, which may lead to longer CPU time and larger proof size (as shown in~\cref{tab:complexity}). We find that setting $m= 4$ yields the best results for both COLE and COLE*.
\end{document}
